\documentstyle[prb,aps,epsf,floats,twocolumn]{revtex}

\begin{document}
\draft

\author{
Hiroki Tsuchiura$^{1}$, Yukio Tanaka$^{2}$, Masao Ogata$^{3}$
and Satoshi Kashiwaya$^{4}$
}
\address{
$^{1}$CREST, Japan Science and Technology Corporation (JST), Nagoya 464-8603,
Japan\\
$^{2}$ Department of Applied Physics, Nagoya University, Nagoya 464-8603,
Japan\\
$^{3}$ Department of Basic Science, Graduate School of Arts and Sciences,
University of Tokyo,\\
 Komaba, Meguro-ku, Tokyo 153-8902, Japan\\
 $^{4}$ Electrotechnical Laboratry, Umezono, Tsukuba, Ibaraki 305-8568,
Japan\\
}
\title{
Local density of states around a magnetic impurity \\
in high-$T_{c}$ superconductors based on the $t$-$J$ model
}
\date{\today}
\sloppy
\maketitle

\begin{abstract}
The local density of states (LDOS) around a magnetic impurity in 
high-$T_{c}$ superconductors is studied using the two-dimensional
$t$-$J$ model with a realistic band structure.
The order parameters are determined in a self-consistent way within the
Gutzwiller approximation and the Bogoliubov-de Gennes theory.
In sharp contrast with the nonmagnetic impurity case, 
the LDOS near the magnetic impurity shows two resonance peaks reflecting
the presence of spin-dependent resonance states.
It is also shown that these resonance states are approximately localized
around the impurity.
The present results have an large implication on the scanning tunneling
spectroscopy observation of
Bi$_{2}$Sr$_{2}$Ca(Cu$_{1-x}$Ni[Zn]$_{x}$)$_{2}$O$_{8+\delta}$.
\end{abstract}

\narrowtext
Due to the recent development in technology, scanning tunneling
spectroscopy (STS) data on high-$T_{c}$ superconductors including
impurities now serve to probe their quasiparticle properties
as well as the nature of the superconducting phase \cite{hudson,yazdani}.
Among various impurities in high-$T_{c}$ superconductors, Zn and Ni 
impurities are fundamental perturbations for the ground states
because they are believed to be substituted for Cu in the CuO$_{2}$ plane
and then strongly disrupt the surrounding electronic structure, especially
the spin configuration.
Thus it is clear that information of quasiparticle states around Zn or
Ni impurities is useful to reveal the mechanism or low-temperature transport
properties of high-$T_{c}$ superconductors.

Quite recently, Pan $et~al.$ succeeded in the STS observation of quasiparticle
resonance states in Bi$_{2}$Sr$_{2}$Ca(Cu$_{1-x}$Zn$_{x}$)$_{2}$O$_{8+\delta}$ 
with high spatial and energy resolution \cite{pan}.
In the vicinity of the Zn impurities, they found an intense quasiparticle
resonance peak at near zero-bias in the STS spectrum.
The corresponding resonance states were observed to be highly localized with 
four-fold symmetry around the Zn impurities. 
Theoretically, the existence of impurity-induced bound states
in $d_{x^{2}-y^{2}}$-wave superconductors
was first predicted by Balatsky $et~al.$ using the $T$-matrix
approximation \cite{bala,bala2,bala3}.
Further, in our previous studies based on the $t$-$J$ model with a
nonmagnetic impurity \cite{tsuchi,lt22}, it was shown that the resonance
states induced by the impurity actually give rise to a resonance peak 
in the local density of states (LDOS)
which can be identified as that observed by Pan $et~al$.
It was also shown that the resonance states are approximately localized
around the impurity.
The obtained spatial dependence of the resonance states can reproduce
the STS result if we take into account the overlapping of the
wavefunctions between STM tip ($s$-wave symmetry) and quasiparticles
($d_{x^{2}-y^{2}}$-wave symmetry) in the CuO$_{2}$
plane \cite{davis}.

Here, we must consider the next question of the quasiparticle 
states around Ni impurities.
The significant difference between Zn and Ni impurities is that,
if they maintain the nominal Cu$^{2+}$ charge, the Zn$^{2+}$ impurity
would have a $(3d)^{10}$, $S=0$ configuration, while Ni$^{2+}$
a $(3d)^{8}$, $S=1$ configuration.
In this framework, the question is how the magnetic moment of the Ni
impurity affects impurity-induced quasiparticle states.
Although there are some theoretical works on quasiparticle states
around a magnetic impurity in $d_{x^{2}-y^{2}}$-wave superconductors
\cite{bala4,poil2,flatte},
no LDOS for direct comparison with STS data on Ni impurities has been
obtained theoretically.
On the other hand, STS observation of Ni-doped BSCCO is only
a matter in time.
Therefore it is urgent to study the LDOS around a Ni impurity in similar
detail to the Zn impurity case \cite{tsuchi}.
In this letter, we study the quasiparticle states around a magnetic impurity
in the same way as in the nonmagnetic (Zn) case.

A Zn impurity in a CuO$_{2}$ plane is considered to be a unitary scatterer
 with $S=0$.
Thus, it has been reasonably modeled by a point-like repulsive potential
\cite{bala,bala2,bala3,tsuchi,lt22,onishi,franz,zhu}
 or a vacant site \cite{poil,fulde} in the previous theoretical works.
On the other hand, behavior of a Ni impurity in a CuO$_{2}$ plane
can be rather complicated.
To take into account the $(3d)^{8}, S=1$ configuration of the Ni impurity
and $d_{x^{2}-y^{2}}$-wave pairing state with a short coherence length in
high-$T_{c}$ superconductors,
we start here with the model studied by Poilblanc $et~al$ \cite{poil2}.
This model is the $t$-$J$ model with a magnetic impurity, which is
allowed to couple to its nearest-neighbor spins via an exchange
coupling $J_{0}$ only
(i.e., the electron transfers onto the impurity site is excluded.).
In order to reproduce the realistic band structure, the next-nearest
neighbor hopping term is added to this model.
Thus the Hamiltonian of the present model is written as
\begin{eqnarray}
{\cal H} &=& -t\sum_{\langle i,j \rangle \neq 0,\sigma}
P_{G}( c^{\dag}_{i\sigma} c_{j\sigma} + {\rm h.c.} )P_{G}
+ J\sum_{\langle i,j\rangle\neq 0}
\mbox{\boldmath $S_{i}\cdot S_{j}$}
\nonumber \\
& & -t'\sum_{ (i,j)\neq 0,\sigma}
P_{G}( c^{\dag}_{i\sigma} c_{j\sigma} + {\rm h.c.} )P_{G}
- \mu\sum_{i,\sigma}c^{\dag}_{i\sigma} c_{i\sigma}
\nonumber \\
& & + J_{0}\sum_{\tau}\mbox{\boldmath $S_{{\rm 0}}\cdot S_{{\rm 0}+\tau}$} 
\label{hamil}
\end{eqnarray}
in the standard notation where $\langle i,j\rangle$ and $(i,j)$ mean the summation over nearest-neighbor and next-nearest-neighbor pairs, 
$\mbox{\boldmath $S_{{\rm 0}}$ }$ represents impurity spin operator with spin-1, and $0+\tau$ represents the nearest-neighbor sites of
the impurity site.
The Gutzwiller's projection operator $P_{G}$ is defined as
$P_{G} = \Pi_{i}( 1 - n_{i\uparrow}n_{i\downarrow})$.
Note that the four bonds connected to the impurity are excluded in the first
three terms.
The last term of Eq.~(\ref{hamil}) corresponds to the coupling of the impurity
to the neighboring spins of the system.
When the impurity coupling $J_{0}$ vanishes, Eq.~(\ref{hamil}) reduces to the
vacancy model which simulates a Zn impurity in a CuO$_{2}$
plane.
Throughout this letter, we take $J/t = 0.2, ~~t'/t = -0.4$ and the hole doping
rate $\delta = 0.15$.
This set of model parameters corresponds to BSCCO near optimal doping
\cite{pan}.

Although Eq.~(\ref{hamil}) is a simplified model for CuO$_{2}$ plane
with a Ni impurity, it is difficult to treat even in a mean-field theory.
Therefore, we make a further simplification to this model,
 i.e., the impurity spin $\mbox{\boldmath $S_{{\rm 0}}$ }$ is treated as 
a fixed Ising-like spin:
$\langle S_{0}^{z}\rangle$.
Thus, the last term of Eq.~(\ref{hamil}) reduces to
\begin{equation}
 J_{0}\langle S_{0}^{z}\rangle\sum_{\tau}S_{\tau}^{z}
 \equiv h_{\rm eff}\sum_{\tau}S_{\tau}^{z}.
\end{equation}
Now the effect of the magnetism of the impurity is represented by the effective
magnetic field $h_{\rm eff} = J_{0}\langle S_{0}^{z}\rangle$
on the nearest-neighbor sites of the impurity.
On the basis of the above simplification, we can proceed to a mean-field
calculation using the Gutzwiller approximation \cite{zhang},
and then we obtain the Bogoliubov-de Gennes (BdG) equation and a set of 
self-consistent equations similar to that in our previous works \cite{tsuchi}:

\begin{equation}
\left(
\begin{array}{cc}
H_{ij}^{\sigma} & F_{ij} \\
F_{ji}^{*} & -H_{ji}^{-\sigma}
\end{array}
\right)
\left(
\begin{array}{c}
u_{j}^{\alpha} \\
v_{j}^{\alpha}
\end{array}
\right)
= E^{\alpha}
\left(
\begin{array}{c}
u_{i}^{\alpha} \\
v_{i}^{\alpha}
\end{array}
\right)
,
\end{equation}

with
\begin{eqnarray}
H_{ij}^{\sigma} &=& -\sum_{\tau}\left( g_{t}t + \frac{3}{4}g_{s}J \xi_{ji}
\right) \delta_{j,i+\tau}(1-\delta_{i,0})(1-\delta_{j,0})   \nonumber \\
& & -\sum_{\nu}g_{t}t' \delta_{j,i+\nu}(1-\delta_{i,0})(1-\delta_{j,0})   \nonumber \\
& & - \left(\mu + \frac{\sigma}{2}h_{\rm eff}\delta_{i,0+\tau}\right)\delta_{ij} 
\nonumber \\
F_{ij}^{*} &=& -\sum_{\tau} \frac{3}{4}g_{s}J(1-\delta_{i,0})(1-\delta_{j,0})
   \Delta_{ij}\delta_{j,i+\tau}  ,
\end{eqnarray}
where $i+\tau$ and $i+\nu$ represent the nearest- and the next-nearest
neighbor sites of the site $i$,
$\sigma = \pm 1$, and $g_{t}$, $g_{s}$ are the renormalization factors
in the Gutzwiller approximation given by
\begin{equation}
g_{t} = \frac{2\delta}{1+\delta},~~g _{s} = \frac{4}{(1+\delta)^{2}}.
\end{equation}
The self-consistent equations are
%
\begin{eqnarray}
\Delta_{ij} &=& \langle c_{i\uparrow}^{\dag}c_{j\downarrow}^{\dag}
\rangle 
=-\frac{1}{4}\sum_{\alpha}
(u_{i}^{\alpha *}v_{j}^{\alpha} +
u_{j}^{\alpha}v_{i}^{\alpha *})\mbox{sgn}(E^{\alpha}) ,
\nonumber \\
\xi_{ij\sigma} &=& \langle c_{i\sigma}^{\dag}c_{j\sigma}
\rangle 
= -\frac{1}{4}\sum_{\alpha}
(u_{i}^{\alpha *}u_{j}^{\alpha} -
v_{j}^{\alpha}v_{i}^{\alpha *})\mbox{sgn}(E^{\alpha}) .
\end{eqnarray}
In the following, we assume $\xi_{ij\uparrow} = \xi_{ij\downarrow}
\equiv \xi_{ij}$ and that $\Delta_{ij}$ is a singlet pairing, i.e.,
$\Delta_{ij} = \Delta_{ji}$.
Since we consider a well-isolated impurity, $\mu$ is fixed to the bulk value  $\mu_{0}$ determined without impurities.


In the present calculation, we regard the $25\times 25$ square lattice as a unit cell of which the impurity is located at the center.
We assume a translational symmetry of $\Delta_{ij}$ with respect to this unit
cell.
We have confirmed that this choice of the size of the unit cell does not have
boundary effects and that the results for an isolated impurity can be
simulated.
Then we make use of the Fourier transform of the BdG equation, for which
we take the number of the unit cells $N_{c} = 20\times 20$.
We solve numerically the BdG equation and carry out an iteration until the self-consistent equations for $\Delta_{ij}$ and $\xi_{ij}$ are satisfied.
Using ($u_{i}^{\alpha}(\mbox{\boldmath $k$})$,
$v_{i}^{\alpha}(\mbox{\boldmath $k$})$) and $E^{\alpha}(\mbox{\boldmath $k$})$,
which are the eigenvectors and eigenvalues of the Fourier transformed BdG equations,
we calculate the LDOS defined by
\begin{eqnarray}
\label{ldos}
N_{i}(E) &=& \frac{1}{N_{c}}\sum_{k,\alpha}\left[ ~|u_{i}^{\alpha}(\mbox{\boldmath $k$})|^{2}~\delta(E^{\alpha}(\mbox{\boldmath $k$})-E) \right. \nonumber \\
&& \left. {}~~~~~~+~|v_{i}^{\alpha}(\mbox{\boldmath$k$})|^{2}~\delta(E^{\alpha}(\mbox{\boldmath $k$})+E) ~\right],
\end{eqnarray}
where $i$ represents a site, 
$\alpha$ is the index of the eigenstates, 
{\boldmath$k$} is the Bloch wave number to the impurity unit cells.

First, we compare the LDOS around the magnetic impurity
with one around the nonmagnetic impurity.
Figure 1 shows that the LDOS obtained on the nearest-neighbor site
for (a) $h_{\rm eff} = 0$ (nonmagnetic) and 
(b) $h_{\rm eff} = 0.16t$.
Note that the dashed lines in Fig. 1 (a) and (b) represent the LDOS obtained on
the site located at the corner of the unit cell, which reproduces the bulk
$d$-wave density of states with V-shaped gap structure.
Here, the superconducting gap edges recover to their bulk value
$E = \pm 0.17t$ for the present parameter choice.
Thus, we can confirm that the impurity is well-isolated in our calculations.
In both nonmagnetic and magnetic cases, we find peaks near the Fermi energy,
reflecting resonances caused by impurity scattering.
The energy levels of the peaks are distinctly different in two cases.
In the nonmagnetic case ($h_{\rm eff}=0$), the resonance peak is found at
slightly {\it below} zero-energy in the LDOS although the impurity scattering
is in the unitary limit.
When $t'=0$, on the contrary, the resonance peak was found
at slightly {\it above} zero-energy \cite{tsuchi,zhu}.
This difference is due to the change of the band structure; here the Van Hove
singularity exists at $E = -0.085t$, i.e., below the Fermi energy.
The ratio of the resonance energy to the energy gap obtained here
($\sim 5\%$) shows a good agreement with the experimental result ($\sim 3\%$)
in Zn-doped BSCCO \cite{pan}.
In the magnetic case, on the other hand, we can see that the resonance peak
splits into two.
\begin{figure}[htb]
\vspace{18pt}
\epsfxsize=6cm
\centerline{\epsfbox{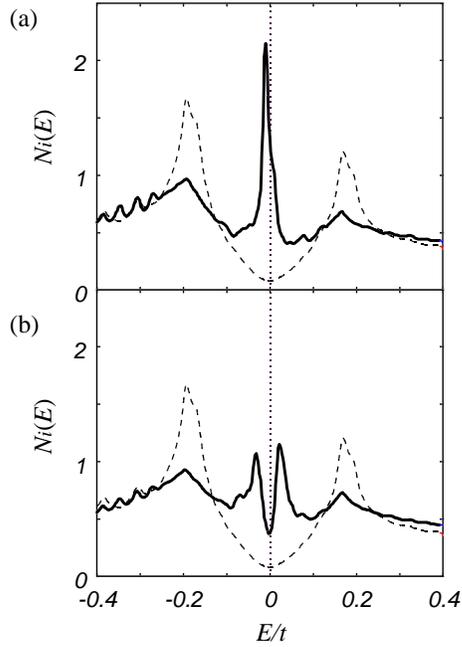}} 
\vspace{0.5cm}
\caption{
The local density of states around the (a) {\it nonmagnetic} and
(b) {\it magnetic} impurity ($h_{\rm eff}=0.16t$).
Solid lines are LDOS obtained on the nearest-neighbor site of the impurity.
Dashed lines represent the LDOS obtained on the site on the corner of the
unit cell.
}
\label{fig:chn}
\end{figure}
%
Since the obtained superconducting order parameters are real, the
peak splitting is not due to the superconducting state with broken 
time-reversal symmetry ( e.g. the $d+is$ state )
but due to the effective magnetic field $h_{\rm eff}$.   
These two peaks correspond to the up and the down spin components of the
quasiparticles.
The peaks are shifted from the resonance energy for the nonmagnetic case because of the energy gain (loss) during the scattering process \cite{kashi}.


It is interesting to investigate the spatial extension of
the impurity-induced resonance states corresponding to the resonance peaks
in Fig. 1 (a) and (b).
In Fig. 2, the DOS with the resonance energies are plotted as a function of
positions around (a) the nonmagnetic and (b),(c) the magnetic impurity.
We can see that these three states are approximately localized around the
impurity (located on the center).
In contrast to the system near half-filling \cite{tsuchi},
the spatial oscillating behavior of these resonance states is not visible
because its period given by the inverse of the Fermi momentum is
incommensurate with $2a$, where $a$ is the lattice spacing.
\begin{figure}[htb]
\vspace{15pt}
\epsfxsize=5.5cm
\centerline{\epsfbox{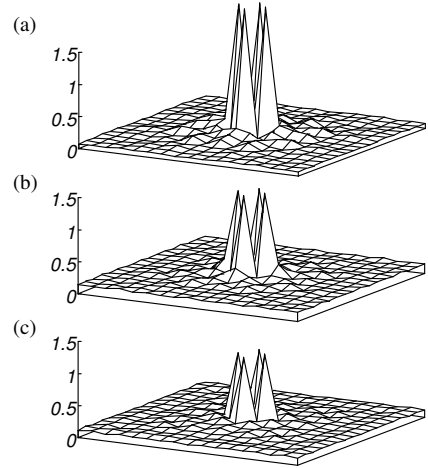}} 
\vspace{0.5cm}
\caption{
Spatial variations of the local density of states at the resonance energy
plotted over $17\times 17$ sites around
 the (a) {\it nonmagnetic} and (b),(c) {\it magnetic} impurity
 ($h_{\rm eff}=0.16t$).
The resonance energies are (a) $E = -0.01t$, (b) $E = 0.022t$ and
(c) $E = -0.032t$.
}
\label{fig:2d}
\end{figure}

Next we study the relation between the amplitude of the peak splitting
and the effective magnetic field $h_{\rm eff}$.
Figure 3(a) shows the LDOS on the nearest-neighbor site of the magnetic impurity for various values of $h_{\rm eff}$.
As $h_{\rm eff}$ is increased, the amplitude of the peak splitting becomes
larger and the two
peaks become broader and smaller.
It is thus confirmed that the splitting is due to the effective
magnetic field.
The $h_{\rm eff}$-dependence of the splitting amplitude $D$ is explicitly
 plotted in Fig. 3(b).
We observe that $D$ is approximately proportional to
$h_{\rm eff}$ when $h_{\rm eff} \leq 0.24t$.

\begin{figure}[htb]
\vspace{5pt}
\epsfxsize=6cm
\centerline{\epsfbox{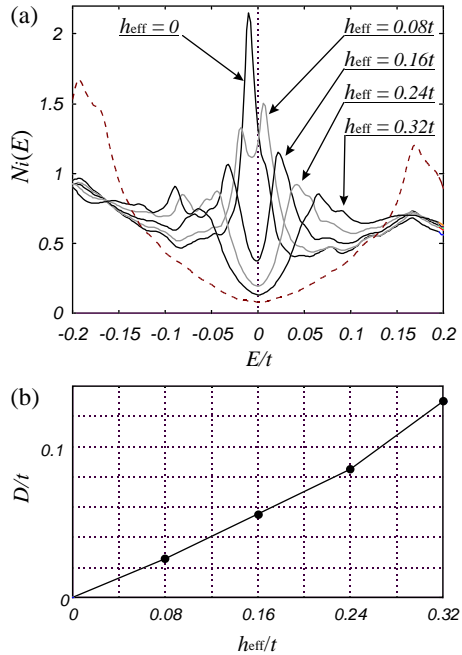}} 
\vspace{0.5cm}
\caption{
(a) The local density of states obtained on the nearest-neighbor site of the
impurity for various values of 
$h_{\rm eff} = J_{0}\langle S_{0}^{z}\rangle$.
The dashed line represents the LDOS obtained on the corner of the unit cell.
(b) The $h_{\rm eff}$ dependence of the amplitude of the peak splitting $D$
obtained on the nearest-neighbor site of the impurity.
}
\label{fig:phase}
\end{figure}

Here, let us discuss whether the peak splitting obtained here can be
actually observed in STS experiment or not.
In order to observe the peak splitting, its amplitude should
be larger than the energy resolution in STS spectra.
If we assume that the present model qualitatively describes the
effects of a Ni spin in a CuO$_{2}$ plane, the amplitude of
$h_{\rm eff}=J_{0}\langle S_{0}^{z}\rangle$ plays a crucial role
in the peak splitting.
However in general, it is difficult to estimate
 $\langle S_{0}^{z}\rangle$ and $J_{0}$.
Recently, in the spin gap state of the $t$-$J$ model, it was shown that the Ni spin (S=1) is partially screened by
 the Cu moments, resulting in an effective impurity spin $S = 1/2$\cite{fulde}.
In that paper, $J_{0}$ is also estimated as $J_{0} = J/2$.
Using these values as a reference, we obtain $h_{\rm eff}\sim 0.25J = 0.05t$
and the expected splitting amplitude is $D \sim 0.015t$ from Fig. 3(b).
If we use $J = 0.13$ eV as a plausible value for CuO$_{2}$ plane,
this estimation gives $D \sim 9.8$ meV.
Of course, we should note that the mean-field calculation of the
$t$-$J$ model may not give a reliable numbers.
However, even taking this point into consideration, we believe that
the value of $D$ is still in order of meV.
In the recent STS experiments, sub-meV resolution in energy
is already accessible \cite{pan}.
Thus the peak splitting would be observed in STS experiments with
high resolution.
In order to make the splitting amplitude $D$ ( ie. $\langle S_{0}^{z}\rangle$)
larger, the observation in an applied magnetic field would be effective.
At this time, it is a natural question whether the splitting observed
under the applied magnetic field is due to the magnetic field itself.
We speculate that the amplitude of Zeeman splitting due to the external
magnetic field of a few Tesla is apparently smaller than that of the peak
splitting obtained here.
Thus, the effect of magnetic impurity will be well identified by checking
the difference in Ni-doped BSCCO and Zn-doped BSCCO under the same magnetic
field.

In our calculation, so far, an antiferromagnetic order parameter and
a $d_{xy}$-wave superconducting order parameter which could be locally
induced around the impurity have not been taken into account.
In particular, it is an interesting problem to reveal whether or not
a $d_{xy}$-wave order parameter with broken time-reversal symmetry is
induced around a magnetic impurity \cite{bala98,movsh}.
However in the standard $t$-$J$ model, a $d_{xy}$-wave order parameter
is not favored, so that some extension of the model is needed.

In summary, we have investigated the LDOS around the magnetic impurity
in the $d_{x^{2}-y^{2}}$-wave superconducting state with short coherence
length based on the $t$-$J$ model.
Within the Bogoliubov-de Gennes theory and the Gutzwiller approximation,
we predict the splitting of the resonance peak in LDOS near the impurity
which is ready to be checked experimentally.

The authors wish to thank J. C. Davis, S. H. Pan, H. Eisaki,
J. Inoue and H. Ito for their useful discussions.
This work is supported in part by a Grant-in-Aid for Creative Basic Research (08NP1201) of the Ministry from Education, Science, Sports and Culture.
Numerical computation in this work was partially carried out at the Yukawa Institute Computer Facility, and the Supercomputer Center, Institute for Solid State
Physics, University of Tokyo.


\end{document}